\documentclass[lettersize,journal]{IEEEtran}
\usepackage{amsmath,amsfonts}
\usepackage{algorithmic}
\usepackage{algorithm}
\usepackage{array}
\usepackage[caption=false,font=normalsize,labelfont=sf,textfont=sf]{subfig}
\usepackage{textcomp}
\usepackage{stfloats}
\usepackage{url}
\usepackage{verbatim}
\usepackage{graphicx}
\usepackage{cite}
\usepackage{booktabs}
\usepackage{microtype} % 微调整字距，减少断行警告
\begin{document}


\title{QSMnet-INR: Single-Orientation Quantitative Susceptibility Mapping via Implicit Neural Representation in k-Space}

\author{
Xuan~Cai, Ruo-Mi~Guo, Xiao-Wen~Luo, Jing~Zhao, Silun~Wang, Tao~Tan, Yue~Liu, Hongbin~Han, and~Mengting~Liu%
\thanks{X. Cai is with the School of Biomedical Engineering, Shenzhen Campus of Sun Yat-sen University, Shenzhen 518107, China (e-mail: caix63@mail2.sysu.edu.cn).}
\thanks{R.-M. Guo and X.-W. Luo are with the Department of Radiology, The Third Affiliated Hospital of Sun Yat-sen University, Guangzhou 510630, China (e-mail: guoruomi86@mail.sysu.edu.cn; luoxw26@mail.sysu.edu.cn).}
\thanks{J. Zhao is with the Department of Radiology, The First Affiliated Hospital, Sun Yat-sen University, Guangzhou 510080, China (e-mail: zhaoj23@mail.sysu.edu.cn).}
\thanks{S. Wang is with Shenzhen Dr. Brain Technologies Co., Ltd.}
\thanks{T. Tan and Y. Liu are with the Faculty of Applied Sciences, Macao Polytechnic University, Macao, China.}
\thanks{H. Han is with (a) the Institute of Medical Technology, Peking University Health Science Center, Beijing 100191, China, and (b) the Beijing Key Laboratory of Magnetic Resonance Imaging Devices and Technology, Peking University Third Hospital, Beijing 100191, China (e-mail: hanhongbin@bjmu.edu.cn).}
\thanks{M. Liu is with the School of Biomedical Engineering, Shenzhen Campus of Sun Yat-sen University, Shenzhen 518107, China (e-mail: liumt55@mail.sysu.edu.cn).}
}

% \markboth{IEEE Transactions on Pattern Analysis and Machine Intelligence,~Vol.~XX, No.~XX, Month~Year}%
% {Cai \MakeLowercase{\textit{et al.}}: QSMnet-INR: Single-Orientation Quantitative Susceptibility Mapping via Implicit Neural Representation in k-Space}

\maketitle

\begin{abstract}
Quantitative Susceptibility Mapping (QSM) quantifies tissue magnetic susceptibility from magnetic-resonance phase data and plays a crucial role in brain microstructure imaging, iron-deposition assessment, and neurological-disease research. However, single-orientation QSM inversion remains highly ill-posed because the dipole kernel exhibits a cone-null region in the Fourier domain, leading to streaking artifacts and structural loss. To overcome this limitation, we propose QSMnet-INR, a deep, physics-informed framework that integrates an Implicit Neural Representation (INR) into the k-space domain. The INR module continuously models multi-directional dipole responses and explicitly completes the cone-null region, while a frequency-domain residual-weighted Dipole Loss enforces physical consistency. The overall network combines a 3D U-Net–based QSMnet backbone with the INR module through alternating optimization for end-to-end joint training. Experiments on the 2016 QSM Reconstruction Challenge, a multi-orientation GRE dataset, and both in-house and public single-orientation clinical data demonstrate that QSMnet-INR consistently outperforms conventional and recent deep-learning approaches across multiple quantitative metrics. The proposed framework shows notable advantages in structural recovery within cone-null regions and in artifact suppression. Ablation studies further confirm the complementary contributions of the INR module and Dipole Loss to detail preservation and physical stability. Overall, QSMnet-INR effectively alleviates the ill-posedness of single-orientation QSM without requiring multi-orientation acquisition, achieving high accuracy, robustness, and strong cross-scenario generalization—highlighting its potential for clinical translation. 
\end{abstract}


\section{Introduction}
\IEEEPARstart{Q}{uantitative} Susceptibility Mapping (QSM) is a key magnetic resonance imaging (MRI) technique for reconstructing tissue magnetic susceptibility distributions from phase data. As an intrinsic physical property of biological tissue, magnetic susceptibility reflects the concentration and spatial distribution of iron, myelin, and calcium. It thus provides valuable insights for investigating neurodegenerative diseases, cerebrovascular abnormalities, and developmental disorders \cite{HaradaKudo-19,BilgicCostagli-122,Singh-ReillySatoh-127,HajiSako-129,RavanfarLoi-105,LorioSedlacik-149,TiepoltRullmann-152,ZhangLiu-159}. For example, excessive iron accumulation in the substantia nigra is a recognized imaging biomarker of Parkinson’s disease, and QSM enables noninvasive quantification of regional iron concentration for early diagnosis and progression monitoring \cite{AlkemadedeHollander-104,ZhangNguyen-145,LiWu-150,LiuWei-155,UchidaKan-157,MohammadiGhaderi-161}.

The core challenge of QSM reconstruction lies in the deconvolution process that estimates susceptibility from the local magnetic field. According to Maxwell’s equations, the local field perturbation can be expressed as a three-dimensional convolution between the susceptibility distribution and the dipole kernel. In the Fourier domain, this relationship becomes pointwise multiplication, where the field spectrum equals the product of the susceptibility map and the dipole kernel. However, in the cone-null region of the dipole kernel—approximately $54.7^\circ$ relative to the main magnetic field direction—the frequency response approaches zero ($D(\mathbf{k}){=}0$). Consequently, susceptibility components along these orientations are unobservable in the measured field, rendering the inversion highly ill-posed. Direct inversion amplifies noise and introduces streaking artifacts, particularly in regions with strong susceptibility gradients such as hemorrhagic lesions, thereby degrading image quality and diagnostic reliability.

Multi-orientation acquisition methods, such as COSMOS (Calculation of Susceptibility through Multiple Orientation Sampling) \cite{LiuSpincemaille-28}, can theoretically recover the missing cone-null information by combining data from at least three distinct head orientations \cite{GkotsouliasJäger-32}. However, this results in long scan times and heavy dependence on patient cooperation, limiting clinical feasibility. In addition, clinical MRI scans are often performed at oblique orientations to better match anatomy or reduce artifacts \cite{KiersnowskiKarsa-126}. Without proper correction, these variations can introduce systematic errors in QSM reconstruction \cite{WuLi-146,HanspachBollmann-147,SpincemailleAnderson-160}.

Single-orientation acquisition remains the clinical standard, yet current algorithms struggle to achieve accurate reconstructions along ill-posed directions. To address this challenge, a variety of QSM reconstruction strategies have been proposed \cite{RobertsRomano-128,FisconeRundo-148,BilgicFan-151,VinayagamaniSheelakumari-156,SalmanRamesh-158}. Traditional methods emphasize physical interpretability through electromagnetic modeling and numerical optimization. For instance, COSMOS achieves theoretically unbiased results but requires multiple acquisitions and is constrained by head-rotation feasibility. Susceptibility Tensor Imaging (STI) \cite{Liu-27} extends susceptibility to a tensor representation to characterize anisotropy, particularly in white-matter fibers, but demands at least six orientations, incurring high computational cost and motion sensitivity. Morphology-Enabled Dipole Inversion (MEDI) \cite{LiuLiu-30} stabilizes single-orientation reconstruction by using magnitude-image edges as spatial priors to guide regularization, reducing artifacts and ensuring structural consistency. Nevertheless, it is vulnerable to noise in magnitude data, where blurred or corrupted edges can cause distortions. The Threshold k-space Division (TKD) method \cite{ShmuelideZwart-29} applies a simple threshold to suppress low-frequency noise, offering computational efficiency but failing to compensate for cone-null information, often leading to streaking and detail loss. The maximum Spherical Mean Value (mSMV) algorithm \cite{RobertsRomano-128} reduces background-field shadows without eroding brain tissue, but its performance remains limited. The background-field-removal network BFRnet \cite{ZhuGao-133} shows improved robustness under pathological conditions by preserving lesion-induced local fields and maintaining stability across orientations, yet it relies on simulated data and remains challenged by noise adaptation and computational cost.

Recent advances in deep learning have accelerated QSM reconstruction by learning complex nonlinear mappings between the magnetic field and susceptibility \cite{BollmannRasmussen-6, LiChen-66}. These methods can be broadly categorized as data-driven or model-driven. Data-driven approaches depend on large labeled datasets to learn direct end-to-end mappings. QSMnet \cite{YoonGong-24}, based on a 3-D U-Net with multi-loss optimization, enables fast reconstruction but generalizes poorly to unseen pathological data and does not explicitly address the cone-null problem, leaving artifact suppression to implicit learning. QSMnet+ \cite{JungYoon-8} improves robustness to out-of-distribution data via data augmentation and linear-response optimization, yet it still struggles with anisotropic regions such as white matter in single-orientation cases. QSMGAN \cite{ChenJakary-26} employs a generative-adversarial framework with enlarged receptive fields to enhance realism and edge sharpness, but training instability and high memory demands limit its practicality. DIAM-CNN \cite{SiGuo-34} separates high- and low-fidelity dipole components using a multi-channel input design, reducing artifacts in hemorrhagic regions, though its threshold parameters require manual tuning and lack adaptively.

Model-driven methods explicitly embed physical priors into neural frameworks to improve interpretability and stability. AutoQSM \cite{WeiCao-22} performs one-step reconstruction without brain extraction, preserving vascular structures and combining numerical optimization with CNN inference to reduce error propagation, but remains prone to boundary artifacts near strongly magnetized regions. MoDL-QSM \cite{FengZhao-23} jointly optimizes network and physical parameters within a susceptibility-tensor model, enhancing physical consistency at the cost of increased complexity. Diffusion-QSM \cite{ZhangLiu-141} achieves superior generalization and detail recovery but suffers from long inference time and discrepancies between patch-based training and full-volume reconstruction. The self-supervised msQSM framework \cite{HePeng-131} demonstrates strong cross-resolution and cross-center generalization, showing promise for neurodegenerative-disease studies, but its robustness to noise and dipole-inversion stability still require improvement. Overall, effectively compensating for missing cone-null information under single-orientation acquisition remains a major bottleneck hindering QSM’s clinical translation.

Implicit Neural Representation (INR) \cite{SitzmannMartel-121} has recently emerged as a powerful paradigm for image reconstruction, 3-D scene modeling, and neural rendering. It employs a multilayer perceptron (MLP) to map continuous spatial coordinates to function values, enabling high-fidelity modeling of continuous fields. Unlike discrete voxel-based representations, INR parameterizes the target field as a continuous function, achieving compact memory usage, fine spatial resolution, and strong interpolation capability \cite{XuMoyer-59}. Zhang et al. \cite{ZhangFeng-18} introduced INR-QSM, which represents susceptibility as a continuous coordinate-based function for single-orientation reconstruction. This unsupervised, subject-specific framework removes the need for labeled data and introduces phase-compensation mechanisms to enhance physical accuracy, partially mitigating deep-learning generalization issues. However, INR-QSM requires per-subject optimization, resulting in low computational efficiency that is unsuitable for clinical deployment. Moreover, its dipole kernel remains fixed and cannot explicitly restore missing cone-null responses in Fourier space. Thus, while INR-QSM established a foundation for the application of INR to QSM, further progress is needed in reconstruction efficiency, structural fidelity, and physical consistency.

In summary, although existing QSM reconstruction methods have made notable progress, they remain limited in modeling the cone-null region, recovering structural details, and maintaining generalization. To address these issues, we propose QSMnet-INR, a framework that combines deep physical modeling with implicit neural representation. The main contributions of this work are as follows.

\subsubsection{Explicit dipole-kernel completion} 
The INR module continuously models the dipole kernel to generate multi-directional dipole responses and compensate for missing information in the cone-null region.

\subsubsection{Self-supervised optimization} 
The INR parameters are updated through back-propagation of QSMnet reconstruction errors, eliminating the need for true dipole-kernel labels and improving cross-dataset and cross-orientation generalization.

\subsubsection{Physics-consistent constraint} 
A frequency-domain residual-weighted Dipole Loss emphasizes error feedback in the cone-null region, ensuring compliance with electromagnetic physical principles.

\subsubsection{End-to-end joint training} 
An alternating optimization strategy enables collaborative updates between the QSMnet backbone and the INR module, enhancing both structural recovery and artifact suppression in single-orientation acquisition.

Through these designs, QSMnet-INR effectively integrates physical modeling and self-supervised learning into an efficient deep framework, providing a new solution for high-quality single-orientation QSM reconstruction with strong stability and promising clinical applicability.

\section{Materials and Method}
\subsection{Framework}
This study proposes QSMnet-INR, a single-orientation QSM reconstruction framework that integrates Implicit Neural Representation (INR) with deep, physics-informed modeling. The key idea is to employ INR to continuously model and complete the traditional dipole kernel, thereby recovering missing information in the cone-null region and mitigating the intrinsic ill-posedness of QSM inversion.

\begin{figure*}[t]
\centering
\includegraphics[width=\linewidth]{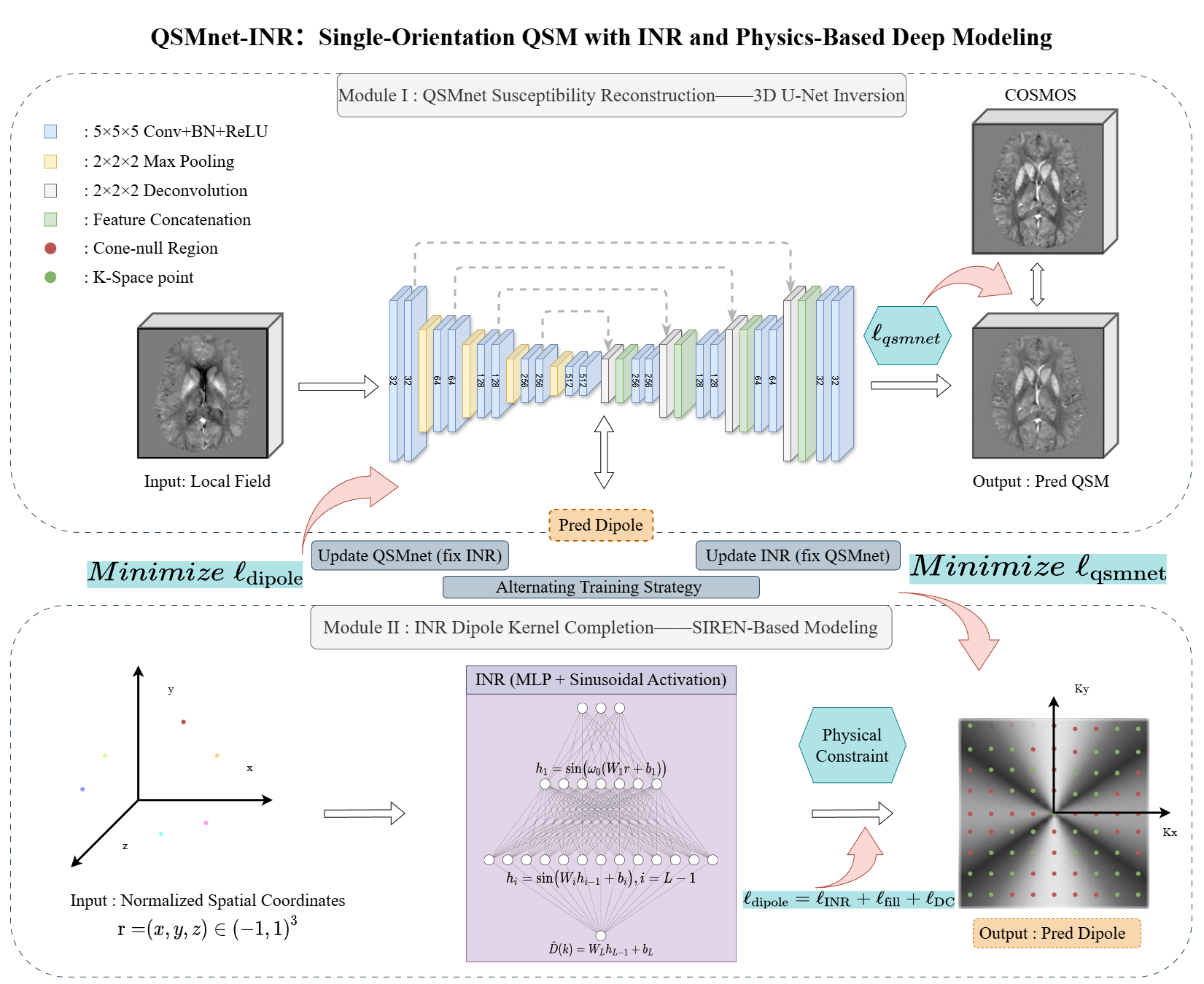}
\caption{Network architecture of the proposed QSMnet-INR framework. 
The framework comprises two cooperative modules: 
(a) QSMnet Susceptibility Reconstruction and (b) INR Dipole Kernel Completion. 
Module~(a) adopts a 3D U-Net architecture that receives the local field as input and generates a predicted susceptibility map $\hat{X}$, supervised by COSMOS ground truth. 
Module~(b), based on a sinusoidal representation network (SIREN), takes normalized spatial coordinates $\mathbf{r} = (x, y, z)$ as input and outputs the predicted dipole kernel $\hat{D}_i(\mathbf{k})$ through a multilayer perceptron (MLP) with sinusoidal activations. 
The two modules are jointly optimized using an alternating training strategy: during the QSMnet update stage, INR parameters are fixed while minimizing $\mathcal{L}_{\text{QSMnet}}$; during the INR update stage, QSMnet parameters are fixed while minimizing $\mathcal{L}_{\text{dipole}} = \mathcal{L}_{\text{INR}} + \mathcal{L}_{\text{fill}} + \mathcal{L}_{\text{DC}}$. 
Each $k$-space point corresponds to a frequency-domain sample, where the cone-null region represents locations with zero dipole response.}
\label{fig_framework}
\end{figure*}

As illustrated in Fig.~\ref{fig_framework}, QSMnet-INR consists of two cooperative submodules: 
\subsubsection{QSMnet Susceptibility Reconstruction Module}
Module I adopts a 3D U-Net architecture that receives the local field as input and produces the predicted susceptibility map (Pred-QSM), supervised by COSMOS results.

\subsubsection{INR Dipole Kernel Completion Module}
Module II, based on a sinusoidal representation network (SIREN) \cite{SitzmannMartel-121}, takes normalized spatial coordinates $\mathbf{r}=(x,y,z)$ as input and outputs the predicted dipole kernel $\hat{D}_i(\mathbf{k})$ through a multilayer perceptron (MLP) with sinusoidal activations.

The two modules are trained jointly using an alternating optimization strategy. During the QSMnet update stage, INR parameters are fixed while minimizing $L_{\text{QSMnet}}$; during the INR update stage, QSMnet parameters are fixed while minimizing $\mathcal{L}_{\text{dipole}}$, which consists of $\mathcal{L}_{\text{INR}}$, $\mathcal{L}_{\text{fill}}$, and $\mathcal{L}_{\text{DC}}$. Each $k$-space point corresponds to a specific frequency-domain sample, and the cone-null region represents points where the dipole kernel equals zero.

\subsection{Cone-Null Response Problem}
For a fixed main magnetic field direction, the dipole kernel in the Fourier domain is defined as
\begin{equation}
D(\mathbf{k}) = \frac{1}{3} - \frac{k_z^2}{|\mathbf{k}|^2}, \quad \mathbf{k} = (k_x, k_y, k_z)
\label{eq_dipole}
\end{equation}
where $|\mathbf{k}|$ denotes the magnitude of the spatial frequency vector $\mathbf{k}$. 
When the angle between $\mathbf{k}$ and the main magnetic field $B_0$ is approximately $54.7^\circ$ (the ``magic angle''), i.e., when $\frac{k_z^2}{|\mathbf{k}|^2} = \frac{1}{3}$, we have $D(\mathbf{k}) = 0$. 
This region forms a cone-null zone in Fourier space, where susceptibility components are unobservable from single-orientation data, leading to unstable inversion, noise amplification, and severe streaking artifacts that degrade image quality.

\subsection{Implicit Neural Representation (INR) Module}
To address this issue, we introduce an implicit neural representation (INR) to continuously model the dipole kernel. 
Specifically, a differentiable function $f_{\boldsymbol{\theta}}$ maps spatial coordinates $\mathbf{r} = (x, y, z)$ to dipole kernel values in frequency space:
\begin{equation}
\hat{D}_i(\mathbf{k}) = f_{\boldsymbol{\theta}}(\mathbf{r}), \quad \mathbf{r} = (x, y, z) \in [-1, 1]^3
\label{eq_inr_mapping}
\end{equation}
The function $f_{\boldsymbol{\theta}}$ is implemented by a multilayer perceptron (MLP) with sinusoidal activations (SIREN structure):
\begin{equation}
\begin{cases}
\mathbf{h}_0 = \mathbf{r}, \\[3pt]
\mathbf{h}_l = \sin\!\big(\omega_0 (\mathbf{W}_l \mathbf{h}_{l-1} + \mathbf{b}_l)\big), \quad l = 1, \dots, L-1, \\[3pt]
f_{\boldsymbol{\theta}}(\mathbf{r}) = \mathbf{W}_L \mathbf{h}_{L-1} + \mathbf{b}_L.
\end{cases}
\label{eq_siren_recursive}
\end{equation}
Here $\omega_0$ is the frequency scaling factor, $L$ denotes the network depth, and $\boldsymbol{\theta} = \{\mathbf{W}_i, \mathbf{b}_i\}$ represents the trainable parameters. 
This design enables accurate modeling of high-frequency local variations. 
Unlike analytical dipole kernels, the INR parameters are updated through backpropagation using QSM reconstruction errors and physical constraints, allowing the network to generate reasonable nonzero responses within the cone-null region in a self-supervised and adaptive manner.

\subsection{Cone-Null Response Loss Function}
In Fourier space, the dipole kernel becomes zero in the cone-null region, causing severe ill-posedness in single-orientation QSM. This instability leads to numerical errors, streaking artifacts, and bias. Traditional methods such as TKD \cite{ShmuelideZwart-29} truncate values below a threshold $t$ in $|D(\mathbf{k})| < t$, preventing numerical explosion but permanently discarding information, resulting in systematic bias. In contrast, DIAM-CNN \cite{SiGuo-34} introduced high- and low-fidelity channels by dividing the frequency domain into high- and low-fidelity parts based on a threshold $\tau$, allowing the network to focus on learning artifact suppression and missing information compensation in the low-fidelity (cone-null) region.

Inspired by these approaches, we propose a cone-null–focused dipole loss that emphasizes both stability and physical plausibility by assigning higher weights to the cone-null region during optimization.

First, a frequency-domain weighting mask is constructed using the analytical dipole kernel $D_{\text{ref}}(\mathbf{k})$:
\begin{equation}
W_{\tau}(\mathbf{k}) = \exp\!\left(-\frac{|D_{\text{ref}}(\mathbf{k})|^2}{\tau^2}\right)
\label{eq_weight}
\end{equation}
where $\tau$ controls the decay from the cone to non-cone regions. 
When $|D_{\text{ref}}(\mathbf{k})|$ approaches zero (inside the cone-null region), $W_{\tau} \approx 1$; otherwise, it quickly decays to zero. 
This weighting amplifies the loss in cone-null regions, forcing the network to focus on optimizing those unstable areas.

To maintain physical consistency, the INR output is regularized by:
\begin{equation}
\mathcal{L}_{\text{INR}} = \sum_{i=1}^{M} \big\| W_{\tau}(\mathbf{k}) \cdot \big(\hat{D}_i(\mathbf{k}) - D_i(\mathbf{k}) \big) \big\|_2^2
\label{eq_linr}
\end{equation}
Unlike TKD’s hard truncation, this soft weighting allows INR to gradually learn effective completion.  
Since zero-response regions differ across orientations, multi-directional kernels can complement each other, as in COSMOS. 
We thus compute the averaged magnitude response:
\begin{equation}
\bar{D}(\mathbf{k}) = \frac{1}{M} \sum_{i=1}^{M} \big| \hat{D}_i(\mathbf{k}) \big|
\label{eq_dbar}
\end{equation}
If $\bar{D}(\mathbf{k}) < \varepsilon$, INR is penalized through:
\begin{equation}
\mathcal{L}_{\text{fill}} = \sum_{\mathbf{k}} W_{\tau}(\mathbf{k}) \cdot \big[\max(0, \varepsilon - \bar{D}(\mathbf{k}))\big]^2
\label{eq_lfill}
\end{equation}
which encourages INR to generate nonzero responses where information is missing.

\begin{figure}[!t] 
\centering 
\includegraphics[width=\columnwidth]{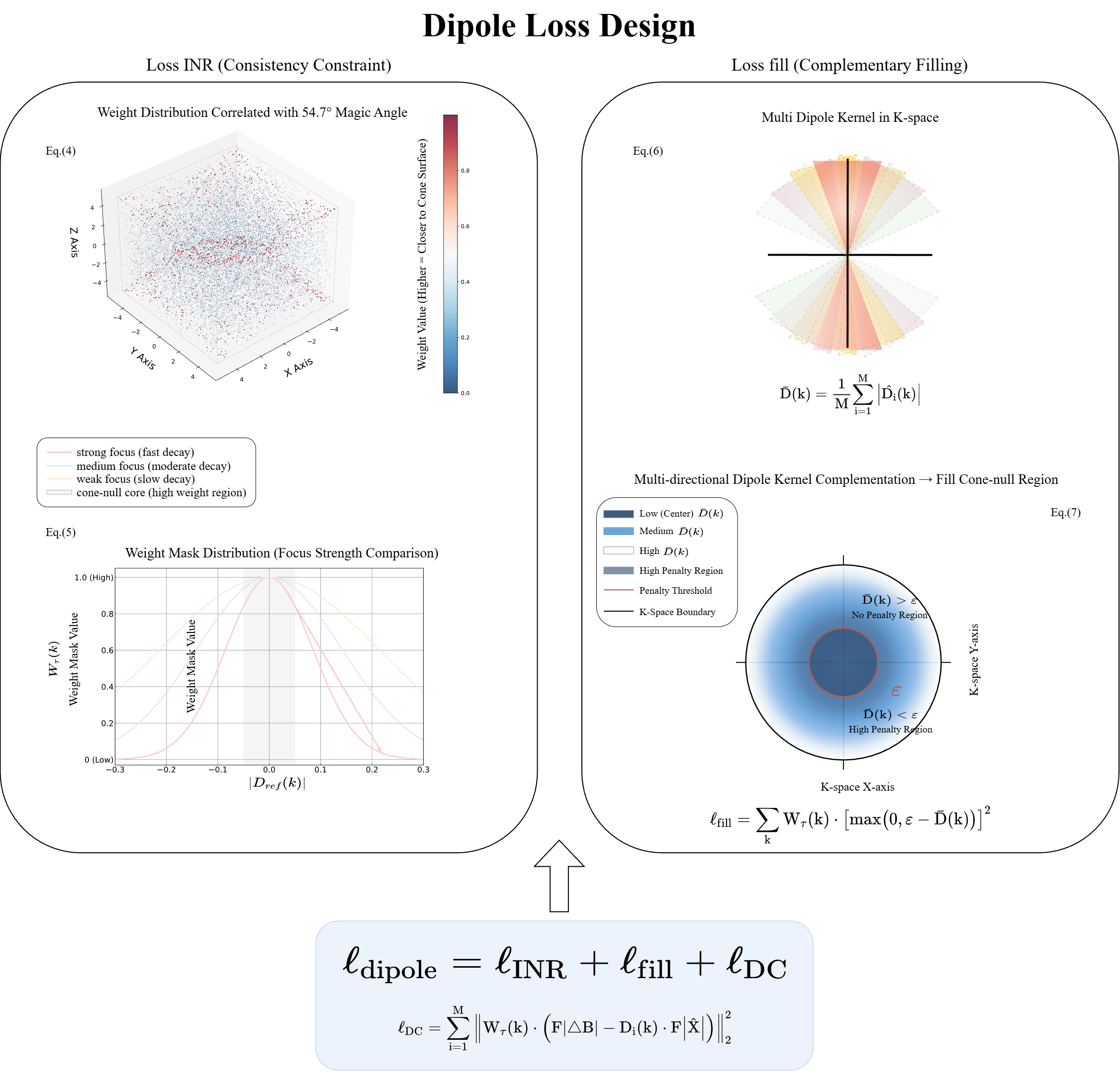}
\caption{Illustration of the cone-null dipole loss function. The loss consists of three parts: INR consistency loss ($\mathcal{L}_{\text{INR}}$), complementary filling loss ($\mathcal{L}_{\text{fill}}$), and data consistency loss ($\mathcal{L}_{\text{DC}}$). The upper-left panel shows the frequency weighting near the $54.7^\circ$ magic angle; points close to the cone receive higher weights. The upper-right panel visualizes multi-directional dipole kernel overlap, while the lower section summarizes the overall formulation $\mathcal{L}_{\text{dipole}} = \mathcal{L}_{\text{INR}} + \mathcal{L}_{\text{fill}} + \mathcal{L}_{\text{DC}}$.} 
\label{fig_INR_Dipole} 
\end{figure}

For data consistency, the predicted susceptibility $\hat{X}$ is constrained to ensure its dipole convolution matches the observed local field $\Delta B$:
\begin{equation}
\mathcal{L}_{\text{DC}} = \sum_{i=1}^{M} \big\| W_{\tau}(\mathbf{k}) \cdot \big(\mathcal{F}\{\Delta B\} - D_i(\mathbf{k}) \cdot \mathcal{F}\{\hat{X}\}\big) \big\|_2^2
\label{eq_ldc}
\end{equation}
where $\mathcal{F}$ denotes the Fourier transform.  
The final dipole loss is thus defined as:
\begin{equation}
\mathcal{L}_{\text{dipole}} = \mathcal{L}_{\text{INR}} + \mathcal{L}_{\text{fill}} + \mathcal{L}_{\text{DC}}
\label{eq_ldipole}
\end{equation}
This formulation enforces physical consistency while guiding INR to generate stable and plausible completions in cone-null regions (see Fig.~\ref{fig_INR_Dipole}).

\subsection{Joint Training Strategy and Total Loss}
In QSMnet-INR, the INR and QSMnet modules are jointly optimized to alleviate the ill-posedness of single-orientation QSM reconstruction. The total loss is defined as:
\begin{equation}
\mathcal{L}_{\text{total}} = \mathcal{L}_{\text{QSMnet}} + \lambda \cdot \mathcal{L}_{\text{dipole}}
\label{eq_total_loss}
\end{equation}
where $\mathcal{L}_{\text{QSMnet}}$ includes model, intensity, and gradient consistency terms to ensure accurate and detailed susceptibility reconstruction, while $\mathcal{L}_{\text{dipole}}$ enforces physically consistent dipole completion in cone-null regions. The parameter $\lambda$ controls the weighting between them.

Training proceeds alternately as follows:
\subsubsection{Update QSMnet}
Fix INR parameters and minimize $mathcal{L}_{\text{QSMnet}}$ to optimize susceptibility reconstruction.
\subsubsection{Update INR}
Fix QSMnet parameters and minimize $mathcal{L}_{\text{total}}$ under self-supervised physical constraints to refine dipole representation.

Through this strategy, INR learns to generate physically plausible and data-adaptive dipole completions without explicit ground truth, thereby improving both the stability and accuracy of single-orientation QSM reconstruction.

\section{Experiments}
\subsection{Experimental Design and Comparative Methods}
To comprehensively evaluate the reconstruction performance of the proposed QSMnet-INR model in quantitative susceptibility mapping (QSM), we conducted systematic experiments on three representative datasets and compared the results with several state-of-the-art QSM reconstruction methods. All experiments were based on single-orientation gradient-echo data. Each method used inputs according to its original design, including local field maps, raw phase maps, or structural priors. Deep learning methods such as QSMnet, DIAM-CNN, and MoDL-QSM typically used background-field–removed local field maps as input, while traditional methods (e.g., TKD, MEDI) and some model-driven approaches utilized raw phase and magnitude images.

The proposed QSMnet-INR framework consists of two cooperative modules: the QSMnet for susceptibility reconstruction and the INR for dipole kernel completion. The QSMnet module takes the local field map as input and is trained in a supervised manner using COSMOS reconstructions as labels. The INR module, in contrast, operates in a fully self-supervised manner—it receives only normalized spatial coordinates as input and updates its dipole kernel parameters through backpropagation of the reconstruction error from QSMnet. This module eliminates the dependence on dipole kernel labels or additional local field maps, providing better flexibility and generalization capability.

To ensure fairness and reproducibility, all methods were executed under a unified preprocessing pipeline and consistent input formats. Parameter settings strictly followed the original publications or official implementations. The comparison covered three representative categories:

\subsubsection{Traditional methods} 
TKD\cite{ShmuelideZwart-29}, MEDI\cite{LiuLiu-30,LiuLiu-3};

\subsubsection{Data-driven deep learning methods}
QSMnet\cite{YoonGong-24}, DIAM-CNN\cite{SiGuo-34}, K-QSM\cite{HeWang-61};

\subsubsection{Model-driven methods}
INR-QSM\cite{ZhangFeng-18}, MoDL-QSM\cite{FengZhao-23}.

\subsection{Datasets}
Three representative datasets were employed to cover standard test scenarios, complex multi-orientation acquisitions, and clinical applications, ensuring good generalization and validation value.
\subsubsection{2016 QSM Reconstruction Challenge Dataset\cite{LangkammerSchweser-111}}
This dataset, released by the Massachusetts General Hospital (Harvard University), consists of single-subject 3T MRI data acquired on a Siemens Tim Trio system using a 32-channel head coil. The acquisition employed rf-spoiled 3D gradient-echo (GRE) sequences with 1.06 mm isotropic resolution, a $240 \times 196 \times 120$ matrix, $\text{TE}/\text{TR} = 25/35$ ms, flip angle $= 15^\circ$, and 15-fold Wave-CAIPI acceleration. A total of twelve head orientations were acquired, enabling COSMOS and STI reconstructions for reference. The dataset provides multi-orientation magnitude and phase images, as well as corresponding anatomical MPRAGE data, allowing comprehensive evaluation of QSM algorithms under standardized conditions. All data are publicly available at \url{http://qsm.neuroimaging.at}.

\subsubsection{Multi-Orientation Gradient-echo MRI Dataset\cite{ShiFeng-13}}
This dataset represents the largest publicly available multi-orientation GRE collection to date, comprising 144 local field maps from eight healthy subjects. All scans were conducted at the Shanghai University of Sport on a 3T Siemens Prisma scanner with a 32-channel head coil. For each subject, over ten head orientations were acquired using a multi-echo 3-D GRE sequence with the following parameters: field of view (FOV) $= 210 \times 224 \times 160$ mm$^3$, matrix size $= 210 \times 224 \times 160$, flip angle $= 20^\circ$, bandwidth $= 190$ Hz/pixel, TR $= 44$ ms, $\text{TE}_1$ / spacing / $\text{TE}_{16}$ $= 7.7 / 5.4 / 38.2$ ms, spatial resolution $= 1$ mm$^3$, and GRAPPA factor $= 2$. All orientations were registered to a reference position using affine transformation to ensure spatial alignment. The dataset provides multi-orientation magnitude and phase images along with COSMOS and STI reconstructions, including six symmetric susceptibility tensor components per subject. The processed data are available at \url{https://osf.io/yfms7/}, and raw GRE data at \url{https://osf.io/y6rc3/}.

\subsubsection{Single-Orientation Clinical Dataset}
This dataset was collected at the Third Affiliated Hospital of Sun Yat-sen University and includes data from ten healthy volunteers and two patients (one with a cerebral hemangioma and one with white-matter abnormalities). The study was approved by the institutional ethics committee, and written informed consent was obtained from all participants. Scans were performed on a GE MEDICAL SYSTEMS scanner with a 48HAP head coil using a 3-D GRE sequence: $\text{TE}_1$ / spacing / $\text{TE}_8 = 3.148 / 2.148 / 18.436$ ms, TR $= 27.412$ ms, bandwidth $= 488.281$ Hz/pixel, flip angle $= 20^\circ$, matrix size $= 256 \times 256 \times 140$, and voxel size $= 1$ mm$^3$.

Additionally, one publicly available intracerebral hemorrhage case was included from Ruijin Hospital (Shanghai, China), acquired on a 3T GE HDxt scanner (matrix size $= 256 \times 256 \times 66$, voxel size $= 0.86 \times 0.86 \times 2$ mm$^3$, TR $= 42.58$ ms, $\text{TE}_1$ / spacing / $\text{TE}_{16} = 3.2 / 2.4 / 39.5$ ms, flip angle $= 12^\circ$) \cite{FengZhao-23}.

All datasets underwent standardized preprocessing—phase unwrapping, background-field removal, registration, normalization, and brain masking—before model input. As single-orientation data lack COSMOS or STI references, evaluations in this setting primarily focused on qualitative assessment to reflect realistic clinical conditions.

\subsection{Evaluation Strategy}
Model performance was assessed through both qualitative visualization and quantitative evaluation. Qualitative analysis involved side-by-side comparisons of predicted susceptibility maps, reference maps, and residuals (Prediction $-$ Label), with particular attention to deep nuclei, tissue boundaries, and fine structural details. Quantitative evaluation employed four widely used metrics—HFEN, NRMSE, SSIM, and PSNR—to assess high-frequency fidelity, intensity accuracy, structural consistency, and overall image quality, respectively.

\subsection{ROI-Based Quantitative Evaluation}
To further investigate the regional performance of different methods, ROI-based quantitative analyses were performed using the 2016 QSM Reconstruction Challenge Dataset and the Harvard–Oxford atlas. Twenty anatomical regions of interest (ROIs) were selected, covering deep nuclei, cortical, and white matter regions. Each ROI’s susceptibility values were extracted and compared with COSMOS results.

\subsubsection{Region selection}
Included structures such as the striatum, globus pallidus, substantia nigra, thalamic nuclei, and cerebellar regions, ensuring coverage of both high- and low-susceptibility areas.

\subsubsection{Quantitative metrics}
The mean, standard deviation, and deviation from COSMOS within each ROI were computed to assess stability and accuracy.

\subsubsection{Visualization}
Radar charts were used to depict the average susceptibility values of each method across 20 ROIs, showing their proximity to COSMOS; box plots displayed residual distributions across brain regions to illustrate robustness.

This ROI-based analysis not only quantified global performance but also revealed local structural differences, providing more interpretable evidence for methodological improvements.

\subsection{Model Analysis Experiments}
To validate the effectiveness of key components in the proposed QSMnet-INR framework and their contributions to susceptibility reconstruction, three sets of model analysis experiments were conducted, focusing on module effectiveness, loss-weight sensitivity, and directional robustness.

\subsubsection{Ablation study}
We analyzed the independent and joint effects of the INR module and the cone-null dipole loss. Three configurations were tested on the 2016 QSM Challenge dataset: a standard QSMnet backbone; QSMnet + INR (without dipole loss constraint); and full QSMnet-INR (with INR and dipole loss). Using identical training strategies and hyperparameters, we compared their HFEN, NRMSE, SSIM, and PSNR to quantify the contribution of each component.

\subsubsection {Loss-weight combination experiment}
To explore the relative impact of each loss term, we decomposed the total loss into three parts: model consistency ($\omega_{\text{model}}$), gradient consistency ($\omega_{\text{grad}}$), and dipole loss ($\omega_{\text{Dipole}}$). By systematically varying their weights and evaluating performance under different configurations, we clarified how each term affects structure preservation, artifact suppression, and cone-null compensation, thereby identifying an optimal balance for practical deployment.

\subsubsection {Direction-sensitivity experiment}
Given the inherent orientation dependency of the dipole kernel, we examined the cross-directional robustness of QSMnet-INR using Sub8 from the Multi-Orientation GRE dataset. The model was trained only on Sub1–Sub7 to avoid data leakage and tested on 18 unseen orientations of Sub8. During inference, network parameters remained fixed while the input direction varied. By analyzing the distributions of HFEN, NRMSE, SSIM, and PSNR across orientations, and examining residuals near cone-null regions, we quantitatively assessed the model’s stability and consistency in cross-orientation reconstruction, validating its generalizability in real-world applications.

\subsection{Experimental Environment}
All training and evaluation were performed on Ubuntu 20.04.2 LTS using Python 3.7 and TensorFlow 1.14. Experiments ran on a workstation equipped with an NVIDIA GeForce RTX 3090 GPU (24 GB VRAM, CUDA 12.4, driver version 550.144.03). Models were trained end-to-end, jointly optimizing the QSMnet backbone and INR module under the unified loss function. All hyperparameters were kept consistent across datasets, and training progress was monitored with TensorBoard. Image-quality metrics were computed using standardized evaluation scripts. To ensure full reproducibility, identical preprocessing, configurations, and pipelines were applied across all datasets. The proposed QSMnet-INR achieved a balanced trade-off between reconstruction accuracy and computational efficiency, supporting stable large-scale training and deployment in clinical susceptibility imaging.

\section{Results}
\subsection{Analysis on the 2016 QSM Reconstruction Challenge Dataset}
\begin{figure}[!t]
\centering
\includegraphics[width=2.5in]{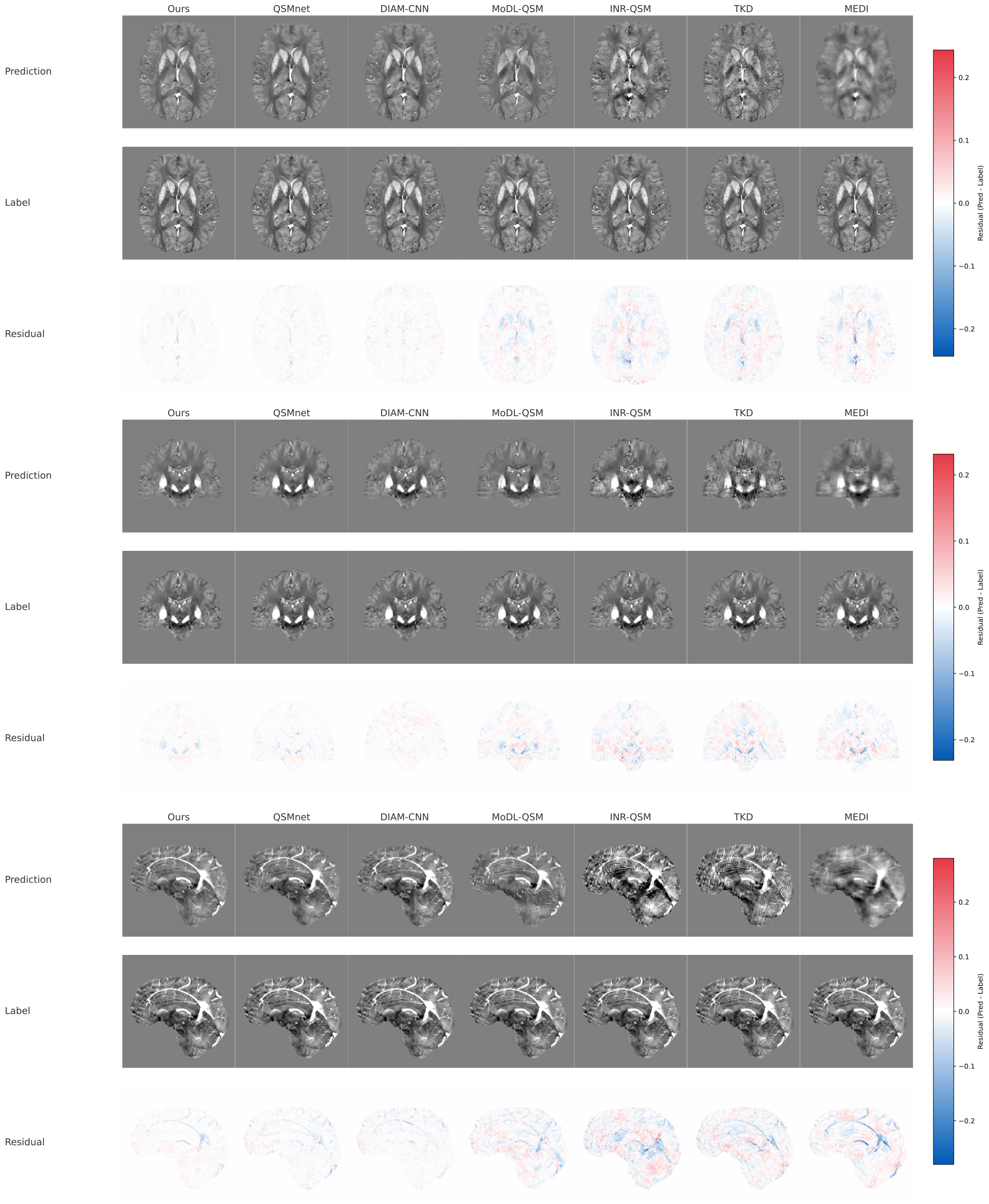}
\caption{Reconstructed susceptibility maps and residuals for different methods on the 2016 QSM Reconstruction Challenge dataset. TKD and MEDI exhibit streaking and blur in high-gradient regions. QSMnet and DIAM-CNN achieve improved structural recovery, while MoDL-QSM and INR-QSM show discontinuities at deep boundaries. QSMnet-INR provides the most accurate and stable reconstruction.}
\label{fig:2016_challenge}
\end{figure}

Fig. \ref{fig:2016_challenge} shows susceptibility reconstructions and residual maps obtained by different methods on the 2016 QSM Reconstruction Challenge dataset. Traditional methods such as TKD and MEDI produce pronounced streaking artifacts and blurred structures in regions with steep susceptibility gradients, including deep gray-matter nuclei and perivenous areas. Their residuals are widespread and of high magnitude, degrading boundary clarity and structural integrity. Deep-learning models QSMnet and DIAM-CNN achieve better overall structural recovery, with DIAM-CNN yielding sharper edges but occasional overfitting. Model-driven methods MoDL-QSM and INR-QSM exhibit local discontinuities at deep-tissue interfaces, where concentrated residuals suggest limited fine-detail modeling.

Among all approaches, QSMnet-INR delivers the highest reconstruction fidelity. It preserves morphological details in cortical boundaries, basal ganglia, and periventricular structures while markedly reducing residual magnitude—only minor errors remain near high-gradient regions. These results confirm stronger structural awareness and more effective artifact suppression, particularly within cone-null areas.

\begin{table}[!t]
\caption{Quantitative comparison on the 2016 QSM Reconstruction Challenge dataset}
\label{tab:results_challenge}
\centering
\footnotesize
\begin{tabular}{lcccc}
\toprule
{Methods} & {HFEN}$\downarrow$ & {NRMSE}$\downarrow$ & {SSIM}$\uparrow$ & {PSNR (dB)}$\uparrow$ \\
\midrule
{Ours} & \textbf{0.362} & \textbf{0.521} & \textbf{0.967} & \textbf{45.669} \\
QSMnet    & 0.591 & 0.891 & 0.898 & 40.999 \\
DIAM-CNN  & 0.500 & 0.676 & 0.946 & 43.399 \\
K-QSM     & 1.684 & 2.111 & 0.669 & 33.509 \\
MoDL-QSM  & 0.691 & 0.692 & 0.834 & 39.753 \\
INR-QSM   & 0.931 & 1.581 & 0.776 & 36.021 \\
TKD       & 0.915 & 1.425 & 0.779 & 36.927 \\
MEDI      & 0.803 & 1.330 & 0.782 & 37.526 \\
\bottomrule
\end{tabular}
\end{table}

Table~\ref{tab:results_challenge} summarizes the quantitative metrics (HFEN, NRMSE, SSIM, and PSNR). QSMnet-INR consistently ranks first across all criteria, achieving the lowest HFEN and NRMSE and the highest SSIM and PSNR, reflecting superior high-frequency fidelity and overall accuracy. In contrast, TKD and MEDI maintain interpretability but show poor numerical precision; QSMnet and DIAM-CNN achieve higher SSIM and PSNR but elevated HFEN, indicating partial loss of fine details. K-QSM and INR-QSM perform worst, with large NRMSE and HFEN values, implying insufficient robustness. Overall, QSMnet-INR attains leading performance on the standard benchmark, excelling in both structural completion and artifact suppression, thereby validating the benefit of integrating physical modeling with implicit neural representations.

\subsection{Analysis on the Multi-Orientation GRE MRI Dataset}
\begin{figure}[!t]
\centering
\includegraphics[width=2.5in]{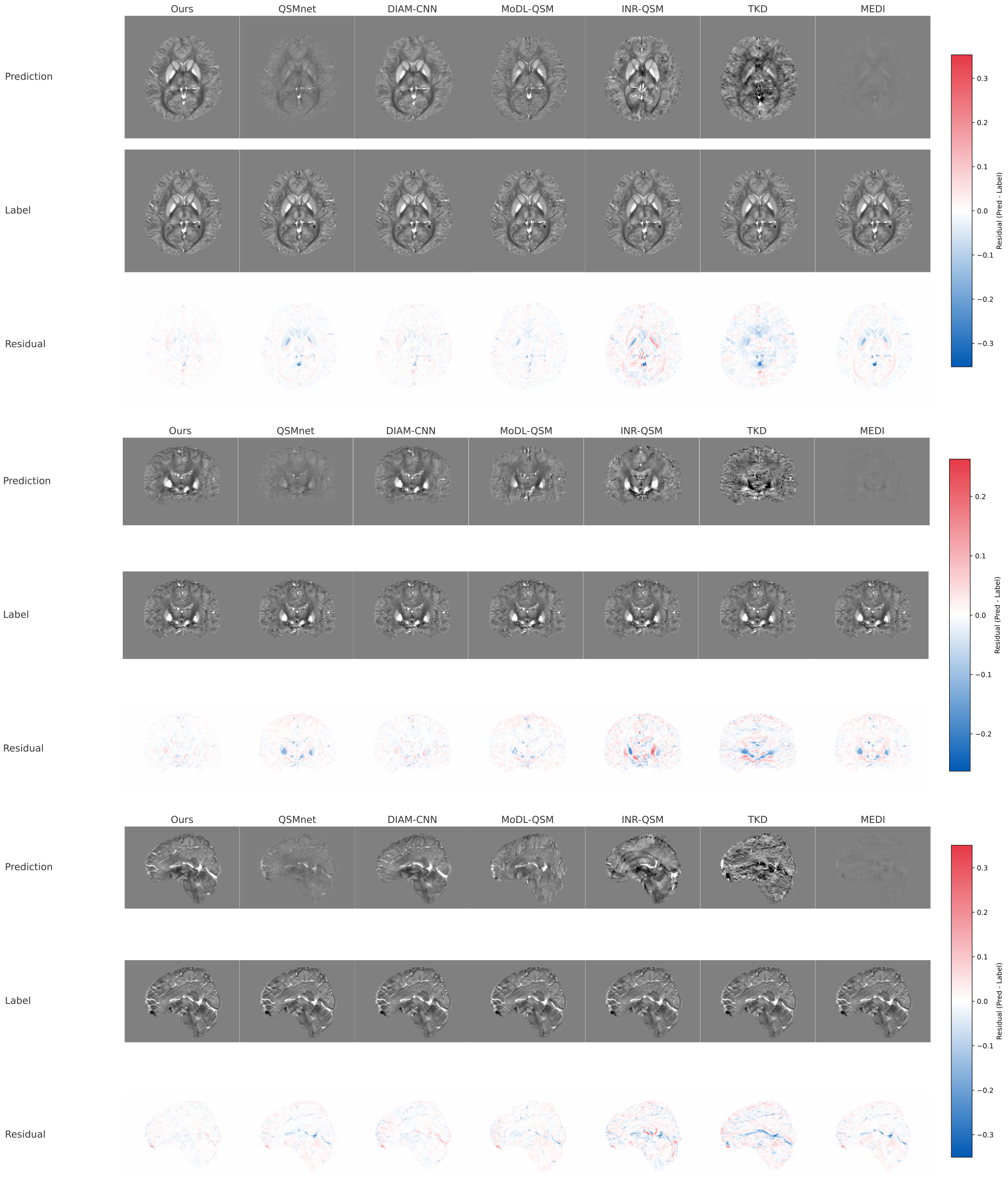}
\caption{Reconstructed susceptibility maps and residuals on the Multi-Orientation GRE MRI dataset. QSMnet-INR best preserves fine structural detail with the lowest residual magnitudes.}
\label{fig:open_data_views}
\end{figure}

To emulate a typical clinical single-orientation scenario, one acquisition direction from the Multi-Orientation GRE dataset was used as input. Fig.~\ref{fig:open_data_views} shows reconstructed susceptibility maps and residuals across methods.

QSMnet-INR accurately restores cortical, striatal, and globus pallidal details, producing morphologically consistent and high-contrast maps with smooth boundaries. In contrast, TKD and MEDI still exhibit pronounced streaking and noise amplification in deep gray matter and perivascular regions. MoDL-QSM and INR-QSM generate overly smoothed patterns in high-frequency areas, while QSMnet and DIAM-CNN better preserve anatomical contours but retain residual artifacts near vessel boundaries and subcortical nuclei. The residual distributions confirm that QSMnet-INR achieves the smallest and most spatially uniform errors, primarily confined to edges or abrupt-transition regions, indicating superior structural fidelity and high-frequency preservation.

Table~\ref{tab:results_multi_orientation} summarizes the quantitative evaluation on the Multi-Orientation GRE MRI dataset. 
QSMnet-INR again achieves the best overall performance, with \textbf{HFEN = 0.493\%}, \textbf{NRMSE = 0.971\%}, \textbf{SSIM = 0.917}, and \textbf{PSNR = 40.25 dB}, 
outperforming all competing methods. Notably, it surpasses DIAM-CNN across all metrics, exhibiting higher SSIM and PSNR and lower HFEN and NRMSE, thereby confirming better detail retention and noise suppression. In contrast, TKD and MEDI show higher HFEN and NRMSE values, reflecting limited robustness and residual streaking artifacts. MoDL-QSM and INR-QSM yield smoother yet less detailed reconstructions, while QSMnet and DIAM-CNN remain competitive but underperform in preserving fine-scale features.

Overall, QSMnet-INR delivers state-of-the-art single-orientation reconstructions by jointly leveraging data-driven detail recovery and INR-based frequency-domain completion, effectively mitigating cone-null ill-posedness and improving reconstruction accuracy and stability.

\begin{table}[!t]
\caption{Quantitative comparison on the Multi-Orientation GRE MRI dataset}
\label{tab:results_multi_orientation}
\centering
\footnotesize
\begin{tabular}{lcccc}
\toprule
{Methods} & {HFEN}$\downarrow$ & {NRMSE}$\downarrow$ & {SSIM}$\uparrow$ & {PSNR (dB)}$\uparrow$ \\
\midrule
{Ours} & \textbf{0.493} & \textbf{0.971} & \textbf{0.917} & \textbf{40.252} \\
QSMnet    & 0.560 & 1.161 & 0.880 & 38.703 \\
DIAM-CNN  & 0.548 & 1.035 & 0.908 & 39.734 \\
K-QSM     & 0.714 & 1.462 & 0.803 & 36.700 \\
MoDL-QSM  & 0.657 & 1.196 & 0.875 & 38.448 \\
INR-QSM   & 1.067 & 2.021 & 0.704 & 33.889 \\
TKD       & 1.024 & 2.071 & 0.698 & 33.675 \\
MEDI      & 0.687 & 1.418 & 0.816 & 36.967 \\
\bottomrule
\end{tabular}
\end{table}

\subsection{Clinical Data Evaluation}
\begin{figure*}[!t]
\centering
\subfloat[]{\includegraphics[width=2.5in]{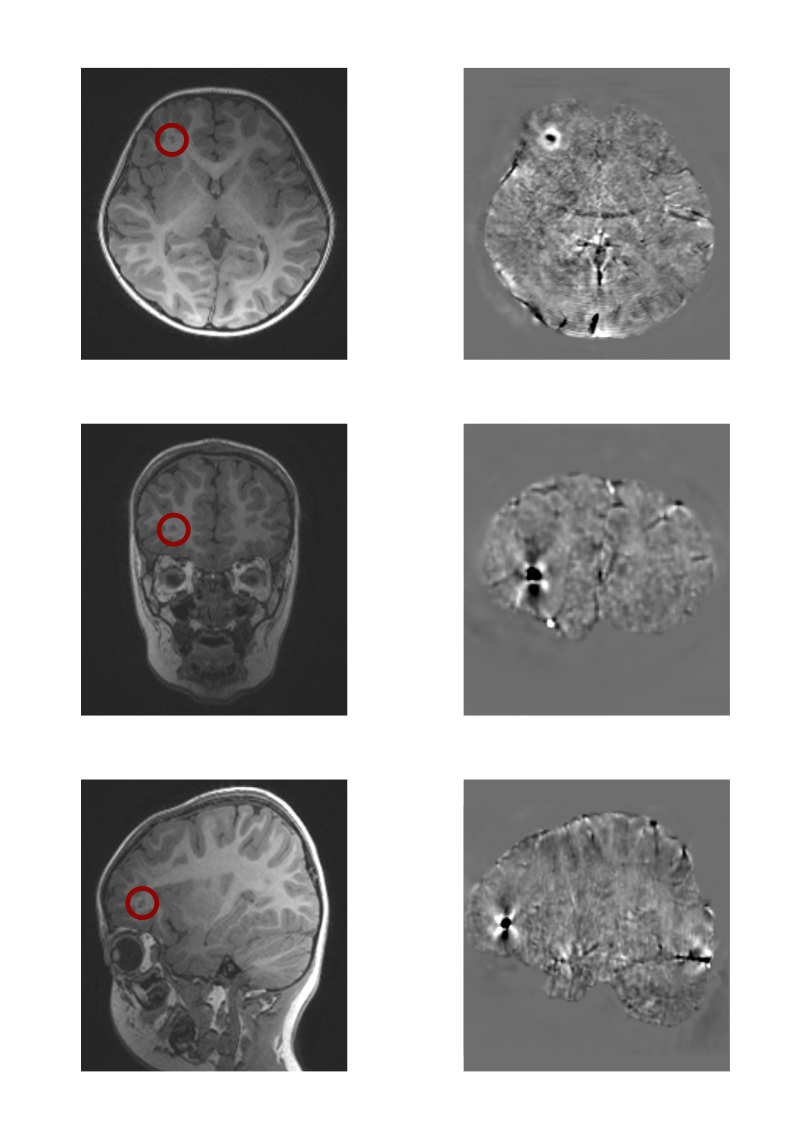}%
\label{fig:hemangioma_case}}
\hfil
\subfloat[]{\includegraphics[width=2.5in]{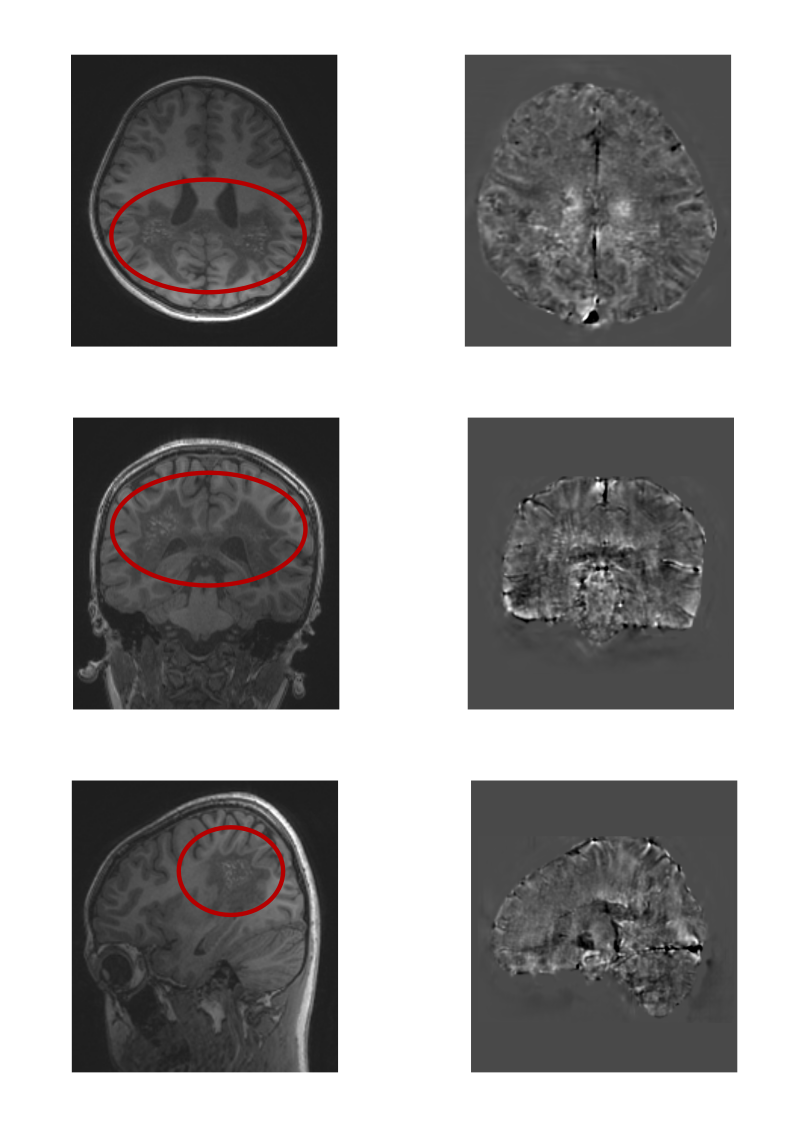}%
\label{fig:wm_case}}
\caption{Clinical QSMnet-INR reconstructions. 
(a) T1-weighted image (left) and QSM reconstruction (right) of a patient with a right-frontal cavernous hemangioma. 
The lesion appears hypointense on T1 and hyperintense on QSM, showing clear boundaries and spatial correspondence. 
(b) T1-weighted image (left) and QSM reconstruction (right) of a patient with abnormal white-matter development. 
The QSM map reveals susceptibility changes consistent with T1 signal abnormalities in bilateral parietal and periventricular regions.}
\label{fig:clinical_cases}
\end{figure*}

Fig.~\ref{fig:clinical_cases} presents representative clinical results reconstructed using QSMnet-INR. 
Subfigure~(a) shows the T1-weighted image (left) and corresponding QSM reconstruction (right) of a patient with a right-frontal cavernous hemangioma. 
The lesion appears hypointense on T1 and markedly hyperintense on the QSM map, consistent with iron-rich blood-product deposition, 
demonstrating QSMnet-INR’s ability to sensitively capture susceptibility alterations with precise spatial correspondence to structural abnormalities. 
Subfigure~(b) depicts results from a patient with abnormal white-matter development. 
The QSM map reveals susceptibility changes aligned with T1 signal abnormalities in bilateral parietal and periventricular regions, 
effectively delineating microstructural white-matter alterations related to pathology.

\begin{figure*}[t]
\centering
\includegraphics[width= \linewidth]{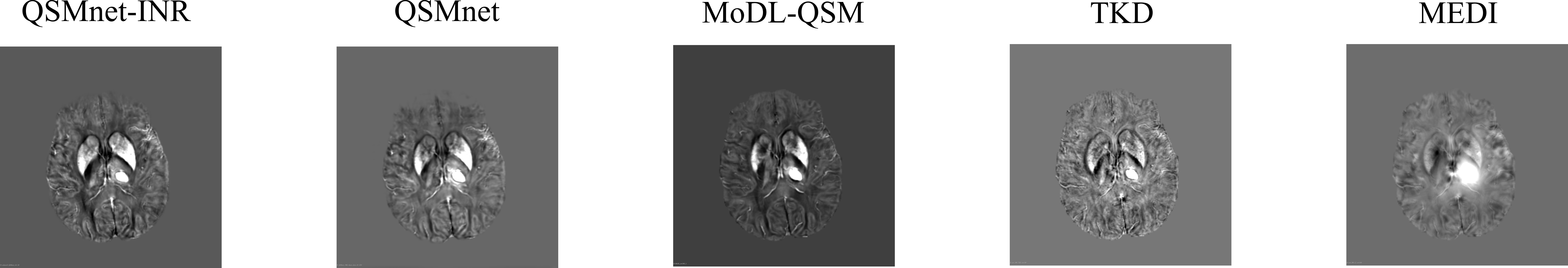}
\caption{Comparison of QSM reconstruction methods in a cerebral hemorrhage case. 
QSMnet-INR, QSMnet, and MoDL-QSM recover the high-susceptibility lesion and boundaries, 
whereas MEDI shows distortion and reduced contrast.}
\label{fig:hemorrhage_case}
\end{figure*}

Fig.~\ref{fig:hemorrhage_case} compares multiple QSM reconstruction methods on a cerebral hemorrhage case~\cite{FengZhao-23}. 
QSMnet-INR, QSMnet, and MoDL-QSM accurately depict the hemorrhagic lesion and its boundaries, 
whereas MEDI exhibits noticeable blurring and contrast loss near the lesion margin.
Overall, these results demonstrate that QSMnet-INR maintains strong lesion visualization and reconstruction stability on clinical single-orientation data, benefiting from its frequency-domain completion mechanism. This highlights its potential for fast and reliable clinical susceptibility imaging.

\subsection{Quantitative Evaluation Across Multiple ROIs}
\begin{figure}[!t]
\centering
\includegraphics[width=\columnwidth]{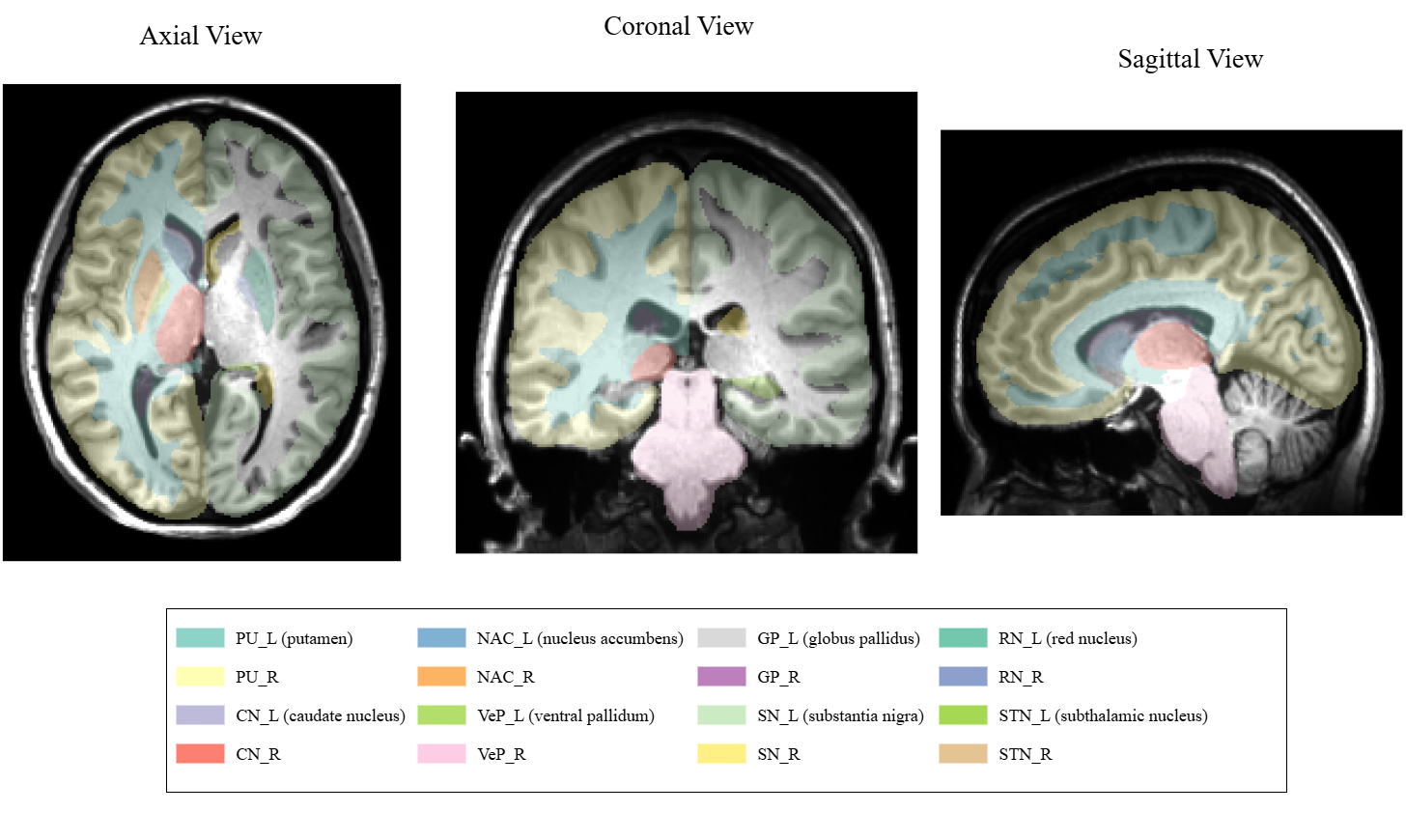}
\caption{Regions of interest (ROIs) overlaid on the T1-weighted image, covering deep gray matter nuclei, cortical, 
and white matter regions used for regional evaluation.}
\label{fig:roi_t1}
\end{figure}

To further evaluate regional reconstruction performance, multiple regions of interest (ROIs) 
were defined on a T1-weighted image, covering deep gray matter, cortical, and white matter structures 
(Fig.~\ref{fig:roi_t1}).

\begin{figure}[!t]
\centering
\includegraphics[width=2.5in]{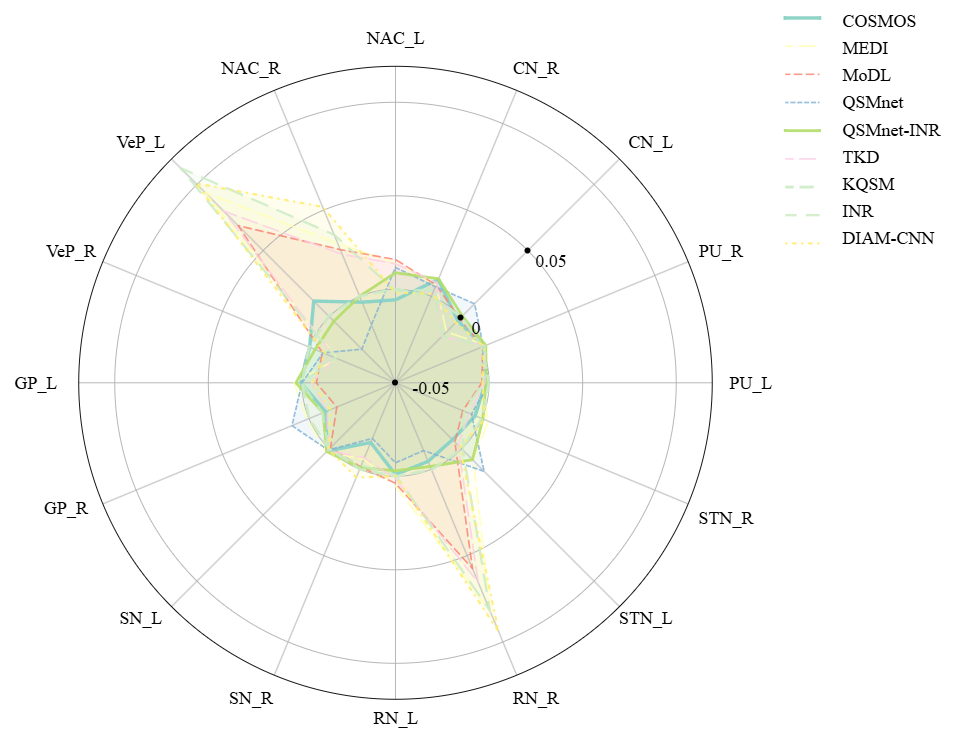}
\caption{Radar analysis across 20 anatomical regions of interest (ROIs). 
QSMnet-INR exhibits the highest overall consistency with the COSMOS reference across most brain regions, 
demonstrating stable regional fidelity and accurate susceptibility quantification in both high- and low-susceptibility areas.}
\label{fig:roi_radar}
\end{figure}

A radar-based regional comparison (Fig.~\ref{fig:roi_radar}) was first conducted to visualize average susceptibility 
values across 20 ROIs relative to the COSMOS reference. 
QSMnet-INR achieves the highest overall alignment with COSMOS across both high-susceptibility regions 
(e.g., globus pallidus, red nucleus, and dentate nucleus) and low-susceptibility regions 
(e.g., white matter tracts and cortical areas), indicating balanced reconstruction accuracy across different tissue types.

\begin{figure*}[!t]
\centering
\subfloat[Signed residuals]{\includegraphics[width=2.5in]{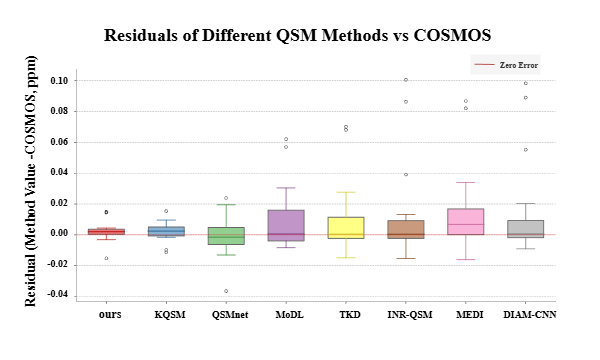}%
\label{fig_residual_signed}}
\hfil
\subfloat[Absolute residuals]{\includegraphics[width=2.5in]{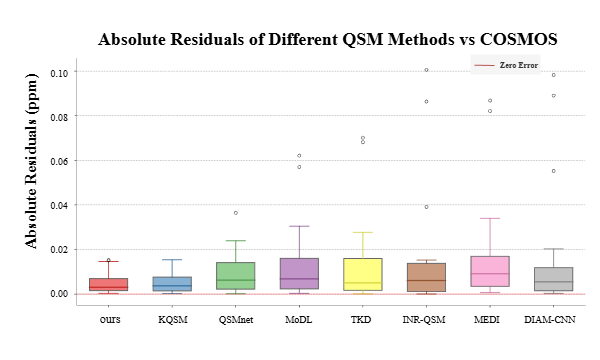}%
\label{fig_residual_abs}}
\caption{Box plot analyses of residuals between different QSM methods and the COSMOS reference across 20 ROIs.
(a)~Signed residuals (Method–COSMOS) reflect bias direction, where QSMnet-INR and K-QSM show compact, near-zero distributions.
(b)~Absolute residuals highlight magnitude stability, showing that QSMnet-INR achieves the lowest regional error and highest robustness across both high- and low-susceptibility regions.}
\label{fig:roi_boxplots}
\end{figure*}

Quantitative evaluation was further performed using residual-based box plot analysis across the same 20 ROIs 
(Fig.~\ref{fig:roi_boxplots}). 
The signed residuals (Method–COSMOS) and absolute residuals were computed to assess both bias and magnitude stability. 
As shown, QSMnet-INR exhibits the most compact and near-zero residual distribution among all methods, 
indicating superior quantitative consistency across brain regions. 
Traditional model-based methods such as TKD and MEDI show large deviations, particularly in high-susceptibility regions 
(e.g., globus pallidus and red nucleus), whereas deep-learning-based methods (QSMnet, MoDL-QSM) 
achieve moderate accuracy but exhibit larger inter-regional variability. 
The absolute residual analysis further confirms that QSMnet-INR preserves susceptibility contrast 
between high- and low-susceptibility tissues more effectively than other methods, 
accurately depicting paramagnetic nuclei and maintaining low errors in diamagnetic white matter. 

Together, these results confirm that QSMnet-INR achieves superior global-to-local fidelity, 
demonstrating robustness across diverse tissue susceptibilities.

\subsection{Model Analysis Experiments}
\subsubsection{Ablation Study}
\begin{figure}[!t]
\centering
\includegraphics[width=2.5in]{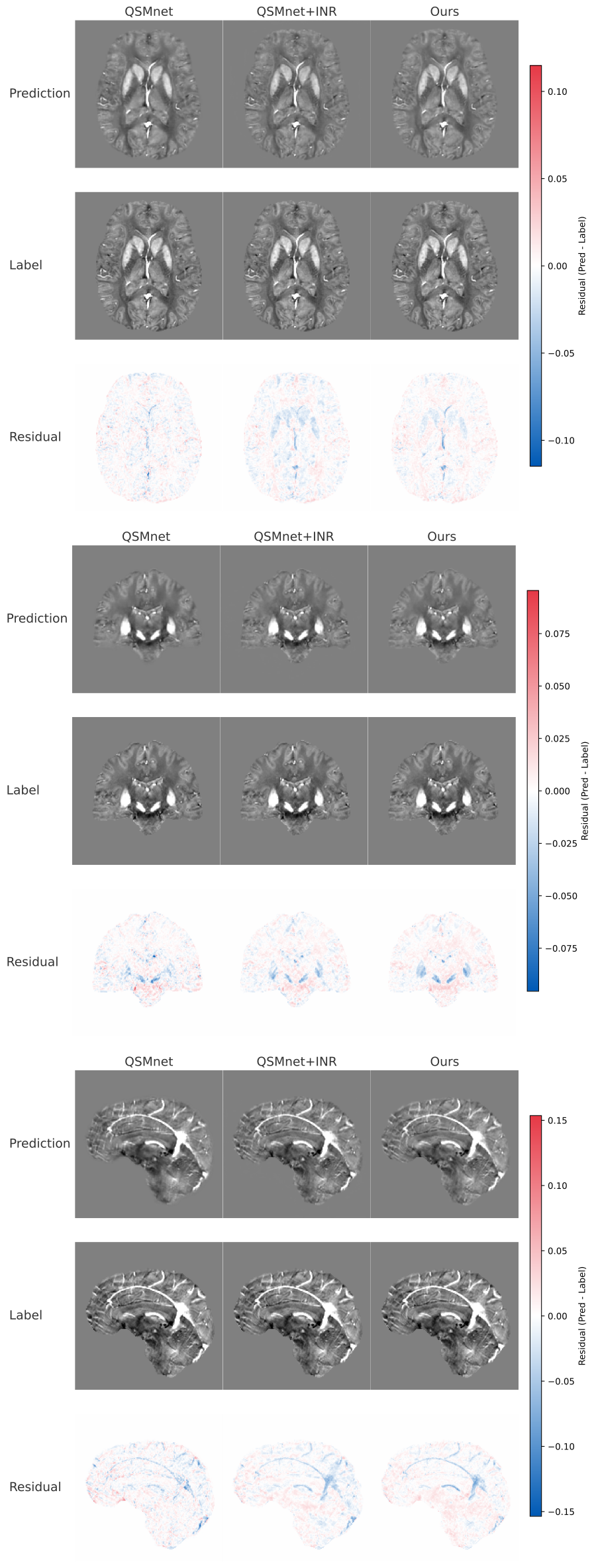}
\caption{Reconstruction results of the ablation study. 
From left to right: QSMnet, QSMnet + INR, and full QSMnet-INR. 
The proposed modules progressively reduce artifacts and enhance structural integrity.}
\label{fig:ablation_study}
\end{figure}

Three configurations were evaluated: QSMnet (baseline), QSMnet + INR, and the full QSMnet-INR. 
As shown in Fig.~\ref{fig:ablation_study} and Table~\ref{tab:ablation_study}, incorporating the INR module markedly reduces 
streaking artifacts and improves structural delineation, demonstrating the benefit of continuous dipole kernel modeling. 
Further inclusion of the cone-null dipole loss yields the best overall performance, achieving balanced improvements across all quantitative metrics. 

These results confirm that INR mitigates information loss by completing the dipole kernel, while the dipole loss enforces frequency-domain regularization; 
their synergy leads to enhanced fidelity and stability in single-orientation QSM reconstruction.

\begin{table}[!t]
\caption{Ablation Study on the 2016 QSM Reconstruction Challenge Dataset}
\label{tab:ablation_study}
\centering
\resizebox{\columnwidth}{!}{%
\begin{tabular}{lcccc}
\toprule
{Model} & {HFEN}$\downarrow$ & {NRMSE}$\downarrow$ & {SSIM}$\uparrow$ & {PSNR (dB)}$\uparrow$ \\
\midrule
QSMnet & 0.591 & 0.891 & 0.898 & 40.999 \\
QSMnet + INR & 0.403 & 0.644 & 0.946 & 43.821 \\
\textbf{QSMnet-INR (Full)} & \textbf{0.362} & \textbf{0.521} & \textbf{0.967} & \textbf{45.669} \\
\bottomrule
\end{tabular}%
}
\end{table}

\subsubsection{Loss-Weight Combination}
To investigate the contribution of different loss components in QSMnet-INR, 
we systematically varied the model consistency weight ($\omega_{\text{model}}$), 
gradient consistency weight ($\omega_{\text{grad}}$), and dipole loss weight ($\omega_{\text{Dipole}}$). 
The quantitative results are summarized in Table~\ref{tab:loss_weight}. 
As shown, $\omega_{\text{model}}$ exhibited the strongest impact on reconstruction quality. 
Moderate weighting improved both structural fidelity and detail preservation, 
while overly large values slightly degraded performance. 
For $\omega_{\text{grad}}$, increasing its value beyond 0.1 led to a decline in all metrics, 
suggesting that excessive gradient regularization may overemphasize edge details at the expense of global accuracy. 
Similarly, the dipole loss weight ($\omega_{\text{Dipole}}$) followed a balanced trend: 
both insufficient and excessive values impaired performance, whereas a moderate weight achieved the best trade-off 
between physical consistency and image quality. 
Overall, the optimal configuration of 
($\omega_{\text{model}}=0.4$, $\omega_{\text{grad}}=0.1$, $\omega_{\text{Dipole}}=0.3$) 
achieved the best overall balance across all evaluation metrics, 
confirming that careful tuning of loss components is critical for stable and high-fidelity QSM reconstruction.

\begin{table}[!t]
\caption{Performance of QSMnet-INR Under Different Loss Weight Combinations}
\label{tab:loss_weight}
\centering
\footnotesize
\begin{tabular}{cccccccc}
\toprule
$\boldsymbol{\omega_{\text{model}}}$ & $\boldsymbol{\omega_{\text{grad}}}$ & $\boldsymbol{\omega_{\text{Dipole}}}$ & {HFEN}$\downarrow$ & {NRMSE}$\downarrow$ & {SSIM}$\uparrow$ & {PSNR (dB)}$\uparrow$ \\
\midrule
0.5 & 0.1 & 0.3 & 0.380 & 0.602 & 0.952 & 44.40 \\
0.6 & 0.1 & 0.3 & 0.355 & 0.538 & 0.963 & 45.39 \\
\textbf{0.4} & \textbf{0.1} & \textbf{0.3} & \textbf{0.362} & \textbf{0.521} & \textbf{0.967} & \textbf{45.67} \\
0.5 & 0.2 & 0.3 & 0.370 & 0.580 & 0.958 & 44.90 \\
0.5 & 0.1 & 0.4 & 0.372 & 0.575 & 0.957 & 44.80 \\
0.5 & 0.1 & 0.2 & 0.358 & 0.540 & 0.962 & 45.90 \\
\bottomrule
\end{tabular}
\end{table}

\subsubsection{Directional Sensitivity Study}
\begin{figure}[!t]
\centering
\includegraphics[width=\columnwidth]{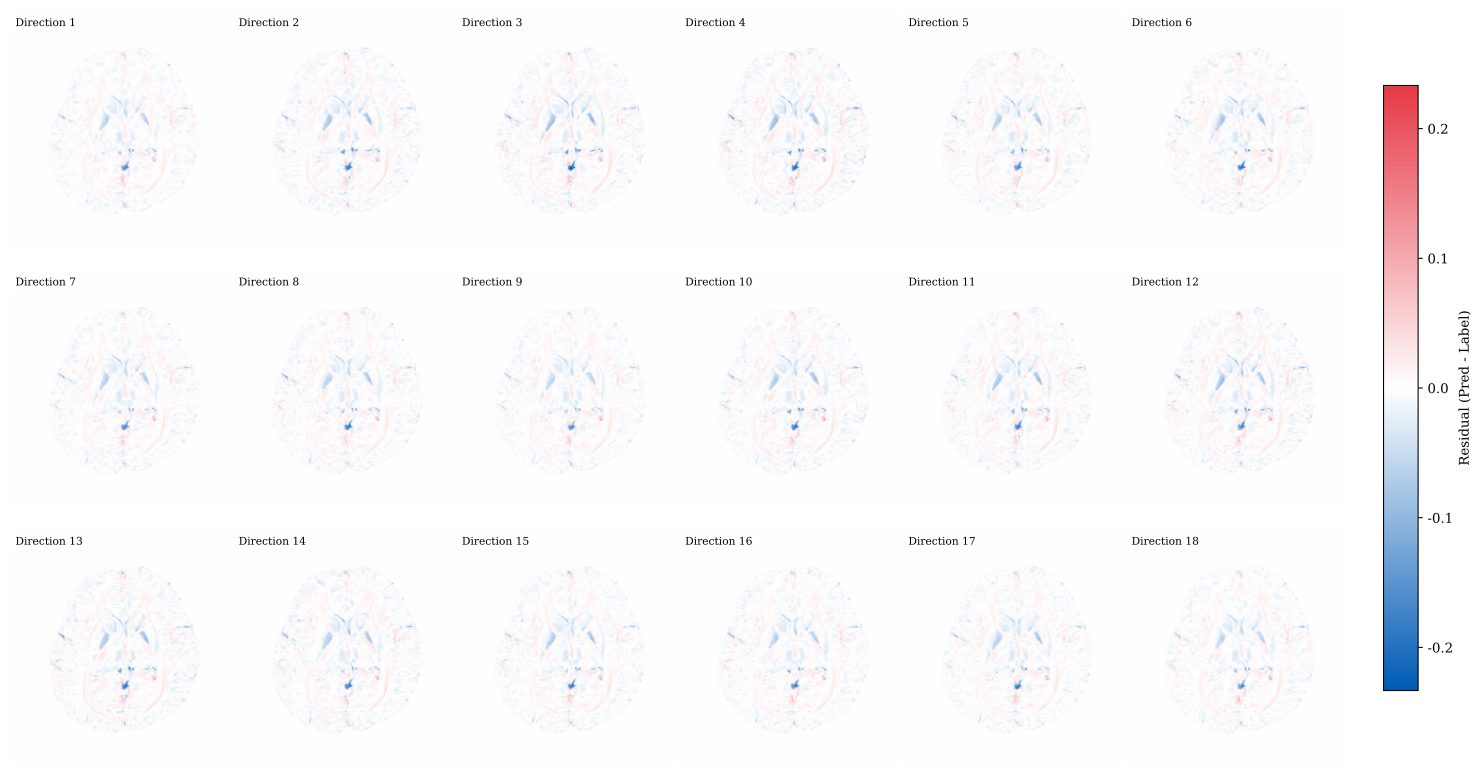}
\caption{Residual maps of QSMnet-INR under 18 single-orientation inputs. 
Residuals remain low across all orientations, mainly localized along structural boundaries and venous regions, 
demonstrating strong reconstruction stability.}
\label{fig:direction_residuals}
\end{figure}

The cone-null region of the dipole kernel is direction-dependent, as it varies with the main magnetic field orientation. 
Consequently, single-orientation QSM reconstructions are sensitive to input direction, especially in areas near the cone-null region. 
To evaluate the directional robustness of QSMnet-INR, we performed a systematic analysis across 18 single-orientation inputs. 
As shown in Fig.~\ref{fig:direction_residuals}, QSMnet-INR maintains consistently low residual magnitudes across all orientations, 
with errors mainly concentrated along structural boundaries and venous regions. 
The model preserves morphological integrity across most directions, reflecting its strong adaptability to orientation-dependent kernel variations.

\begin{figure}[!t]
\centering
\includegraphics[width=\columnwidth]{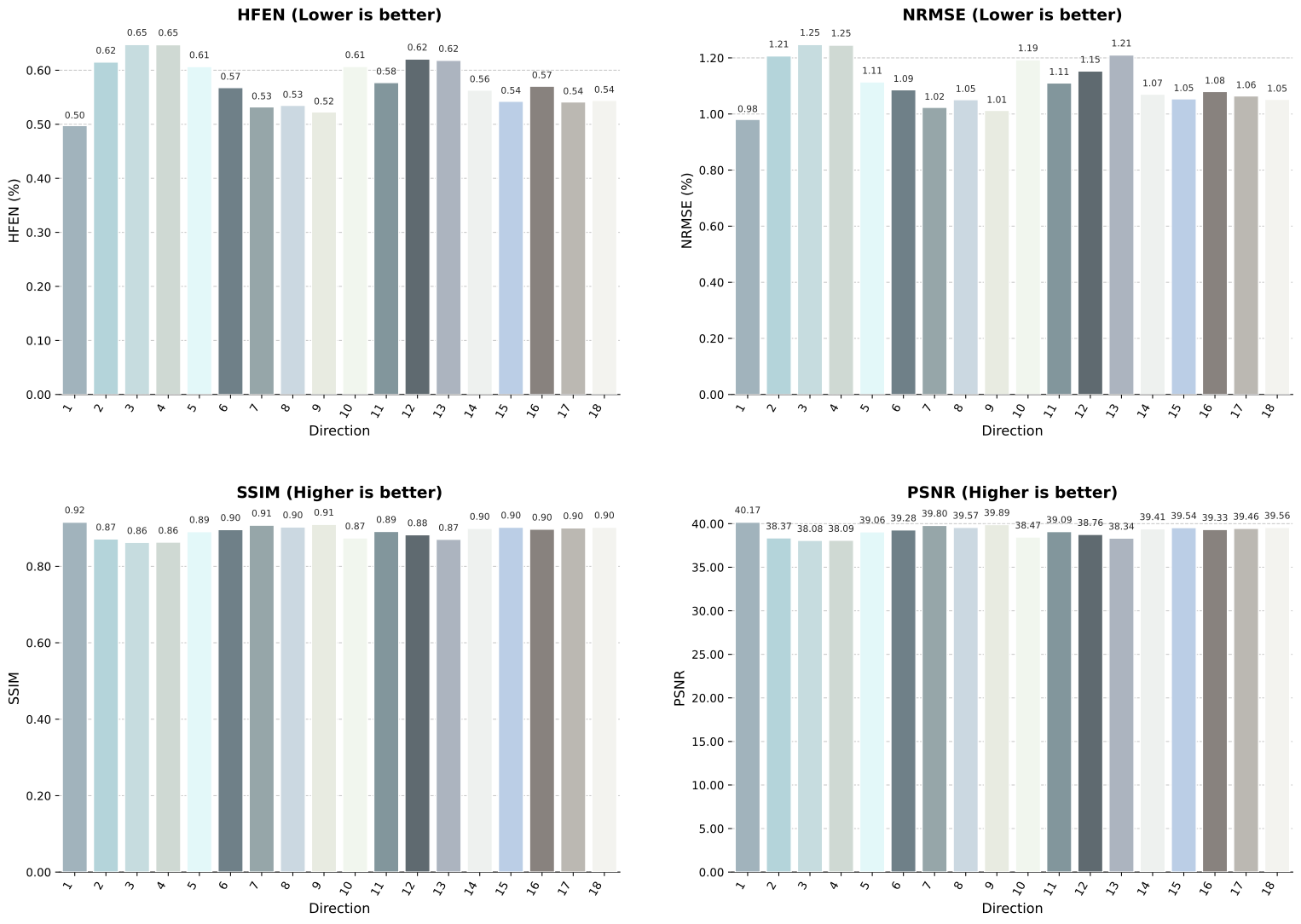}
\caption{Quantitative performance of QSMnet-INR across 18 input orientations. 
Metric fluctuations are minimal (SSIM $>$ 0.88, PSNR $>$ 39~dB in most directions), 
confirming consistent detail preservation and high directional robustness.}
\label{fig:direction_metrics}
\end{figure}

Quantitative evaluation results are summarized in Fig.~\ref{fig:direction_metrics}. 
Across all orientations, the metrics fluctuate within narrow ranges—HFEN and NRMSE variations remain below 0.2 and 0.27, respectively—
indicating stable detail preservation. 
Moreover, in 15 out of 18 orientations, the SSIM exceeds 0.88 and PSNR remains above 39 dB, confirming robust structural and intensity consistency. 

In summary, QSMnet-INR demonstrates strong directional generalization, maintaining stable reconstruction quality and structural fidelity 
under varying input orientations. 
This robustness suggests that the model can effectively handle clinical scenarios with uncontrolled or inconsistent acquisition directions.

\section{Discussion and Conclusion}
This study comprehensively evaluated the proposed QSMnet-INR framework across multiple benchmark datasets and clinical cases, demonstrating its effectiveness in mitigating the ill-posed dipole inversion problem in single-orientation QSM. Through systematic experiments—including ablation, loss-weight, and directional sensitivity analyses—QSMnet-INR consistently exhibited superior reconstruction fidelity, reduced artifacts, and enhanced structural consistency, confirming both the physical rationality and empirical robustness of its design.
\subsection{Methodological Insights}
The core innovation of QSMnet-INR lies in integrating an Implicit Neural Representation (INR) module with a cone-null response–aware loss, bridging frequency-domain physics modeling and spatial-domain deep-learning reconstruction. The INR module continuously completes the dipole kernel within the cone-null region, effectively recovering missing frequency components and alleviating zero-cone artifacts. Meanwhile, the cone-null loss explicitly constrains the ill-posed regions, guiding the network to focus on unstable frequency components during optimization. Together, these mechanisms form a complementary framework that balances data-driven learning and physics-based consistency.

Ablation studies confirm that each component independently improves reconstruction accuracy, while their combination yields the most stable, artifact-free, and physically consistent susceptibility estimates.

\subsection{Reconstruction Performance}
Across all datasets, QSMnet-INR achieved \textbf{state-of-the-art quantitative and qualitative performance}. On the \textbf{2016 QSM Reconstruction Challenge}, it surpassed both traditional and deep-learning-based methods in all major metrics (HFEN, NRMSE, SSIM, PSNR), producing sharp anatomical boundaries and effectively suppressing streaking artifacts. On the \textbf{Multi-Orientation GRE MRI dataset}, QSMnet-INR maintained high reconstruction quality even with single-orientation input, accurately recovering fine cortical and subcortical structures and demonstrating strong generalization. In clinical evaluations, it captured susceptibility variations within pathological regions with improved clarity and contrast, providing superior lesion depiction and diagnostic reliability compared to existing approaches.

Although both benchmark datasets were originally acquired using multiple orientations, only a single orientation was employed in this study to emulate realistic clinical conditions. Under such settings, where the cone-null problem becomes most severe, QSMnet-INR fully exploits its frequency-domain completion capability. As a result, it consistently outperforms DIAM-CNN across all quantitative metrics, achieving higher SSIM and PSNR and lower HFEN and NRMSE. On high-SNR multi-orientation data, where the inversion problem is less ill-posed, the performance gap naturally narrows but remains in favor of QSMnet-INR.

These findings establish QSMnet-INR as particularly advantageous for single-orientation or ill-conditioned scenarios, offering a reliable solution for practical clinical applications.

\subsection{Stability and Robustness}
The loss-weight combination and directional sensitivity analyses further verified the stability of QSMnet-INR. 
Optimal reconstruction quality was achieved with moderate weighting of the model consistency loss 
($\omega_{\text{model}} = 0.4$) and dipole loss ($\omega_{\text{Dipole}} = 0.3$), 
balancing structural fidelity and fine-detail preservation. 
Excessive gradient-weighting led to edge overemphasis and minor global degradation, 
suggesting the importance of balanced training constraints. 
Directional sensitivity experiments confirmed that QSMnet-INR maintained stable performance across 
18 distinct input orientations, with minimal fluctuations in HFEN and NRMSE and consistently high SSIM values. 
These results demonstrate strong robustness to variations in magnetic field orientation and acquisition geometry, 
a crucial property for clinical applicability.

\subsection{Clinical Significance and Future Perspectives}
Beyond methodological and experimental validation, QSMnet-INR holds strong potential for clinical translation. 
Quantitative Susceptibility Mapping (QSM) is a noninvasive biomarker sensitive to iron content, myelin concentration, 
and hemoglobin derivatives—key indicators for neurological disorders such as Parkinson’s disease, Alzheimer’s disease, 
and multiple sclerosis. 
However, conventional single-orientation QSM remains limited by its ill-posed inversion and artifact sensitivity. 
By introducing a frequency-domain completion mechanism, QSMnet-INR improves reconstruction reliability, 
offering a feasible pathway for rapid and artifact-resistant single-orientation QSM in clinical practice.

The framework is also scalable and extensible. 
Its implicit representation strategy could be adapted to other MRI modalities affected by incomplete or noisy data, 
such as diffusion tensor imaging (DTI) or functional MRI (fMRI). 
Future work may explore cross-modal constraints (e.g., incorporating T1/T2 priors), 
multi-task learning, or lightweight architectures to enhance clinical interpretability and computational efficiency.

Although the INR module slightly increases training and inference time, 
techniques such as model compression, parallelization, or knowledge distillation could further improve scalability. 
Moreover, large-scale validation on multi-center clinical cohorts will be essential to fully establish generalizability. 
From a physical modeling standpoint, the current framework assumes isotropic susceptibility, 
which may not fully capture anisotropic effects in myelinated white matter. 
Integrating Susceptibility Tensor Imaging (STI) and multi-echo data fusion could further enhance accuracy in anisotropic tissues~\cite{MengAllen-132}. 
Additionally, extending the framework to disentangle paramagnetic and diamagnetic sources—for instance, 
by incorporating relaxation rate ($R_2$, $R_2'$) information—may provide deeper insights into iron, myelin, and calcium interactions 
in neuropathology~\cite{LiFeng-14, KanUchida-15, GuevaraRoche-125}.

\subsection{Summary}
In summary, \textbf{QSMnet-INR} represents a significant methodological and practical advance in QSM reconstruction. It achieves robust cone-null completion, consistent performance across orientations, and superior artifact suppression while maintaining physical interpretability. he integration of implicit representation learning with physics-informed constraints ensures both scientific soundness and clinical relevance.

With continued optimization and large-scale clinical validation, QSMnet-INR holds strong potential to become a reliable tool for quantitative susceptibility imaging, enabling precise brain iron mapping, lesion detection, and improved understanding of neurodegenerative diseases.

\bibliographystyle{IEEEtran}
\bibliography{refs}

\vfill

\end{document}